\documentclass[useAMS,usenatbib,singlespacing]{mn2e}
\usepackage{graphicx,amssymb}
\title[Cosmic Ray Acceleration in Flux Tubes]
{Cosmic Ray Acceleration in  Hydromagnetic Flux Tubes}
\author[AR Bell, JH Matthews, KM Blundell, AT Araudo ]
{AR Bell$^{1}$\thanks{E-mail:Tony.Bell@physics.ox.ac.uk},
JH Matthews$^{2}$,  KM Blundell$^{2}$, and AT Araudo$^{3,4}$ \\
$^{1}$University of Oxford, Clarendon Laboratory, Parks Road, 
Oxford OX1 3PU, UK\\
$^{2}$University of Oxford, Astrophysics, Keble Road, 
Oxford OX1 3RH, UK\\
$^{3}$Astronomical Institute of the Czech Academy of Sciences, Bocni II 1401, CZ-14100 Prague, Czech Republic\\
$^{4}$ELI Beamlines, Institute of Physics, Czech Academy of Sciences, 
25241 Doln{\'\i} B\v re\v zany, Czech Republic \\
}
\begin{document}
\date{}
\pagerange{\pageref{firstpage}--\pageref{lastpage}} \pubyear{2019}
\maketitle
\label{firstpage}
\begin{abstract}
We find that hydromagnetic flux tubes in back-flows in the lobes of radio galaxies
offer a suitable environment for the acceleration of cosmic rays (CR)
to ultra-high energies.
We show that CR can reach the Hillas (1984) energy even if  the magnetised turbulence in the flux tube 
is not sufficiently strong for Bohm diffusion to apply.
First-order Fermi acceleration by successive weak shocks in a hydromagnetic flux tube  is shown to be equivalent to second-order Fermi acceleration
by strong turbulence.
\end{abstract}
\section{ INTRODUCTION}
The origin of ultra-high energy cosmic rays (UHECR)  is uncertain,
and many possible sources have been proposed.
Radio galaxies have long been considered a likely source of UHECR because of their high power, large size and longevity
(eg Hardcastle et al 2009, O'Sullivan et al 2009, Wykes et al 2013, Eichmann et al 2018).
In Matthews et al (2018, 2019)
we made the case from observations and theory that UHECR may be accelerated by
shocks in the lobes of radio galaxies.
Our numerical simulations demonstrated the presence of shocks in
concentrated, approximately annular, flows centred on the jet axis, which we refer to as hydromagnetic flux tubes, 
or more simply as `flux tubes',
that emerge from the high pressure hot-spots at the end of relativistic jets.
Back-flows in radio lobes have been discussed by many authors 
(for example Norman et al 1982, Falle 1991, Scheuer 1995, Saxton et al 2002,
Krause 2005, Keppens et al 2008, Mignone et al 2010, Cielo et al 2014, Tchekhovskoy \& Bromberg 2016).
The flow expands into the lobes where the pressure is lower.
As discussed in Matthews et al (2019), from application of the Bernoulli equation, the flow becomes supersonic
before being slowed down in one or more shocks as it progresses deeper into the lobe
as depicted in an idealised model in figure 1.
Our simulations of radio jets  support this picture by demonstrating the
presence of a succession of shocks in flux tubes with Mach numbers of a few and
flow velocities of the order of $c/4$.
With these velocities, shocks are 
well suited to UHECR acceleration since fully relativistic shocks 
pose severe problems for acceleration to the highest energies (Kirk \& Reville 2010; Lemoine \& Pelletier 2010; Reville \& Bell 2014; Bell et al 2018).
Further discussion of our model and the observational context can be found in Matthews et al (2018, 2019).
Simulations show that  flux tubes are usually more contorted than the idealised flux tubes in figure 1.
However, the flux tube need only be long enough to contain the CR diffusion scaleheights upstream and downstream of a shock,
and curvature should not invalidate the model since magnetic field lines threading the tube should cause CR
to diffuse predominantly parallel to the tube.
\begin{figure}
\includegraphics[angle=0,width=6cm]{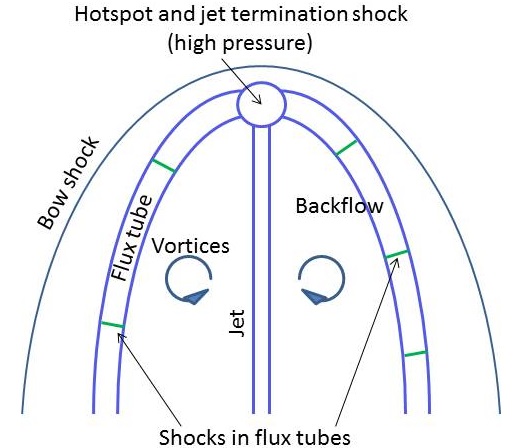}
\centering
\caption{
Idealised model of shocks and flux tubes in the lobes of radio galaxies.  
}
\label{fig:figure1}
\end{figure}
Cosmic ray (CR) acceleration by diffusive shock acceleration 
(Krymskii 1977, Axford et al 1977, Bell 1978a,b, Blandford \& Ostriker 1978)
has been studied at length, both theoretically and observationally,
in the context of supernova remnants (SNR).
The maximum energy of CR accelerated by SNR is constrained by various factors.
The most general constraint is set by the physical size $R$ of the shock,
leading to a maximum CR energy of  $ZuBR$ (Hillas 1984), where $u$ is the flow velocity, $B$ is the magnetic field,
and the CR has charge $Ze$.
$ZuBR$ is the characteristic maximum CR energy in eV if $u$, $B$ \& $R$ are in SI units. 
A related constraint is set by the need for the shock to persist for a time longer than the time taken for CR to be accelerated
(Lagage \& Cesarsky 1983a,b).
For a quasi-parallel shock (magnetic field parallel to or at an angle significantly  smaller than $\pi/2$ to the shock normal)
the Lagage \& Cesarsky limit is equivalent to the Hillas limit if the shock persists for a flow time $R/u$.
For perpendicular shocks, the acceleration is very rapid and the maximum CR energy is set by the Hillas limit  (Jokipii 1982,1987).
In this paper we assume that the hydrodynamic flow through a flux tube is in steady state on the scale of the time taken for CR acceleration, 
in which case the limit on the maximum CR energy is better understood through the spatial limit of Hillas than the timescale analysis of Lagage \& Cesarsky.

The characteristics of backflows in relation to Bernoulli's equation are discussed by Matthews et al (2019)
where it is found that relatively narrow flux tubes are surrounded by vortices
as idealised in figure 1.
Flux tubes have two characteristic scale-lengths, namely the length and width of the flux tube, and it is not immediately clear
which of these, if either, should be invoked when applying the Hillas limit.

In the case of SNR, 
CR acceleration can be limited by the time taken to
amplify the magnetic field (Zirakashvili \& Ptuskin 2008, Bell et al 2013), but this limitation 
is less stringent in flux tubes since (i) the backflow passes through a succession of shocks such that amplification at one shock prepares
a magnetic field for subsequent shocks, 
(ii) a large magnetic field is generated as the plasma passes from the jet through the termination hot-spot into the backflow
(Araudo et al 2015, 2018).
Hence the situation is different from that of a single isolated shock at the outer boundary of a SNR.
A different analysis is needed to determine the maximum CR energy attained in a steady long-lived back-flow.
Here we construct an idealised model designed to reproduce the dominant features relevant to CR acceleration
by shocks in flux tubes in radio galaxies.

In section 2 we consider first-order acceleration at a single stationary shock in a non-relativistic back-flow.
Acceleration at shocks with finite spatial extent has been considered previously in the context of the Earth's bow shock
(Eichler 1981, Lee 1982, Webb et al 1985).  Eichler's analysis exhibits many of the features we find to apply to flux tubes,
most notably that the maximum CR energy is independent of the strength of the turbulent scattering of CR by fluctuations in the magnetic field.

In section 3, we apply second-order Fermi theory (Fermi 1949, Jones 1994) to acceleration by strong turbulence in a flux tube and demonstrate the
equivalence between this and first-order acceleration by successive weak shocks.

Our analysis focusses on the application of our model to the lobes of radio galaxies, but
there may be other scenarios in which the model might apply such as
particle acceleration at shocks inside jets with a variable speed, shocks in jets encountering stars or clouds,
or reconfinement shocks in jets.
Depending on the circumstances, 
the model may need to be extended to oblique shocks or relativistic shocks.

We use SI units throughout with CR energies in eV.
We use the symbol $T$ for CR energies for consistency with related papers where we wish to avoid possible confusion with the electric field.
UHECR may consist of nuclei with various charges and masses.
To avoid unnecessary mathematical complication, we consider only the acceleration of protons.
The behaviour of heavier nuclei can be read from our equations by noting that if a proton with charge $e$  is accelerated
 to ultra-high energy $T$ by the processes considered here,
 then a nucleus with charge $Ze$ is automatically accelerated to energy $ZT$ by exactly the same processes.

\section{ ACCELERATION AT A SINGLE STATIONARY  SHOCK IN A FLUX TUBE}
\begin{figure}
\includegraphics[angle=0,width=8cm]{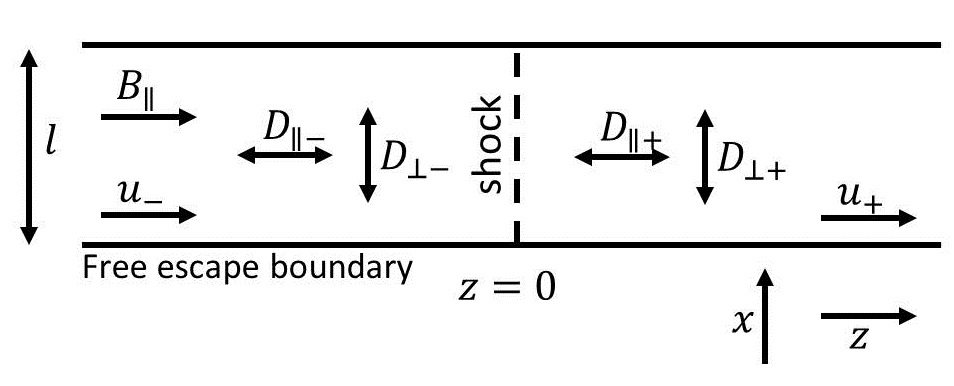}
\centering
\caption{
Flux tube geometry with a single stationary shock.
The $D{\rm s}$ are diffusion coefficients with subscripts $\parallel$ or $\perp$ denoting diffusion directed parallel or perpendicular to the flow.
The subscripts $-$ or $+$ denote the upstream or the downstream of the shock respectively.
}
\label{fig:figure2}
\end{figure}
 Our model for UHECR acceleration at a single shock in a back-flow is visualised in figure 2.
We assume 2D geometry in which the flux tube, which might  be more accurately referred to as a flux slab, is infinite and uniform in the $y$ direction.
Slab geometry is closer  to the actual geometry in a radio lobe
where the flow geometry resembles in cross-section an annulus extending in angle around the jet, as may be imagined from figure 1
and as seen in Matthews et al (2019).
The flux tube contains a stationary shock at $z=0$ where $z$ is the distance along the flux tube. 
The flow is from left to right in $z$.
The flux tube of width $l$ occupies the space $-l/2<x<l/2$ across the tube.
The flux tube is assumed to be straight in $z$ on the basis that it bends over distances larger than diffusion  lengths characterising the process of  particle acceleration.
The plasma flows at velocity $u(z)$ with a velocity $u_-$ into the shock, and exits the shock with  a downstream velocity $u_+$.

We assume that there is a uniform magnetic field aligned along the flow direction, $B=B_\parallel$ both upstream and downstream of the shock.
This parallel magnetic field is made plausible by the assumption that all the plasma in the tube has 
recently passed through the termination shock,
 so it is magnetically disconnected from plasma in the lobe outside the tube.
The crucial consequence in our model is that CR propagating along a magnetic field line stay inside the flux tube 
and can only escape from the sides of the flux tube by diffusing across the magnetic field.
Moreover, the plasma in the flux tube is stretched in the $z$ direction as it expands out of the hot spot to become supersonic as it passes through the lobe,
thus making the field predominantly locally
\onecolumn\noindent
\noindent
parallel to the flow rather than perpendicular across the width of the tube. 
As discussed below, CR diffusion has different 
diffusion coefficients
$D_\parallel $ and $D_ \perp $ parallel and perpendicular to the magnetic field.
Overall, the model depicted in 
figure 2 is a simplification but it elucidates basic principles that could be expanded upon in further work.

The equation for the evolution of the CR distribution function $f$ is (eg Blandford \& Eichler 1987)
$$
\frac {\partial f}{\partial t} 
+\frac {\partial (fu)}{\partial z} 
-\frac {\partial }{\partial z} \left ( D_\parallel \frac {\partial f}{\partial z} \right )
-\frac {\partial }{\partial x} \left ( D_\perp  \frac {\partial f}{\partial x} \right )
- \frac {1}{3} \frac {\partial u}{\partial z} \frac{1}{p^2} \frac{\partial (p^3 f)}{\partial p}
=0
\eqno {(1)}$$
where $p$ is the magnitude of momentum, and $f$ is defined in the local rest frame of a plasma moving at velocity $u(z)$.
In this section we assume that $u$, $D_\perp $ \& $D_\parallel$ are spatially uniform except that they jump
discontinuously in value across the shock.
Separation of variables gives a steady state solution ($\partial f/\partial t = 0$) of the form
$$
f_\pm (z,x,p)=\sum _{m=0}^\infty f_{sm} (p) \exp (k_{m\pm} z) \cos \left (\frac {(2m+1) \pi x}{l} \right )
\eqno{(2)}$$
where $f_-$ and $f_+$ are the solutions upstream and downstream respectively.
$f_{sm}(p)$ is the value of the $m$th Fourier component of both $f_-$ and $f_+$ at the shock ($z=0$)
since  $f$ is continuous across the shock: $f_s(x,p)=f_-(0,x,p)=f_+(0,x,p)$.
The boundary condition at the edges of the flux tube is that $f=0$ on the assumption that CR reaching the boundary escape freely.
$\partial u /\partial z =0$ everywhere except at $z=0$.

As shown in the Appendix (equation A6), the time-independent solution of equation (1) is
$$
f_{sm}(p)=F_{m} p^{-\frac {3u_ -}{u_--u_+}} 
\left [\frac {1}{2}+  \sqrt { \frac {1}{4}+\frac { k_m ^2 D_{\parallel -}D_{\perp -} } {u_-^2} }\ \right ] 
^{ {3u_-}/{2(u_--u_+)}}
\left [\frac {1}{2}+  \sqrt { \frac {1}{4}+\frac { k_m ^2D_{\parallel +}D_{\perp +}  } {u_+^2} }\ \right ] 
^{{3u_+}/{2(u_--u_+)}}
\times
$$
$$
\exp \left [ \frac {3}{2} \frac {u_- + u_+}{u_- - u_+}  \right ]
\exp \left [
-  \frac{3}{2} 
\frac {  
\sqrt { u_-^2 + 4 k_m ^2 D_{\parallel -}D_{\perp -}   } +  \sqrt { u_+^2 + 4 k_m ^2 D_{\parallel +}D_{\perp +}  } 
  }{u_--u_+}  \right ]
\eqno{(3)}$$
\noindent
where  $F_m$ are constants of integration and $k_m=(2m+1) \pi /l$. 
From equation (1), the rate at which the shock accelerates CR to momenta greater than $p$ is
$$
\frac {\partial N}{\partial t}=
\int _{-\infty} ^ {\infty} dz
\int _{-l/2} ^ {l/2} dx
\int _{p} ^ {\infty} 4 \pi p^2 dp
\left [
\frac{1}{3} \frac {\partial u}{\partial z} \frac{1}{p^2} \frac{\partial (p^3 f)}{\partial p}
\right ]
=\frac {4 \pi}{3} (u_- - u_+) p^3 \int _{-l/2}^{l/2} f_s dx
$$
$$
{\rm giving} \hskip 1 cm
\frac {\partial N}{\partial t}=
\frac {8 (u_- - u_+)l }{3}
\sum_{m=0}^ \infty \frac {(-1)^{m}}{2m+1} p^3 f_{sm} .
\eqno{(4)}
$$
The resulting differential energy spectrum is $ n(p)=-dN/dp$.
Differentiating equation (4), and  using equation (A4) for $\partial (p^3 f_{sm})/ \partial p$,
the rate at which CR are produced with momentum $p$ is
$$
\frac {\partial n}{\partial t}=
\sum_{m=0}^ \infty \frac {4l(-1)^{m}}{2m+1}
\left [ 
 \sqrt {u_-  ^2 +4 k_m^2  D_{\parallel -} D_{\perp -} } - u_- + \sqrt {u_+  ^2 +4 k_m^2  D_{\parallel +} D_{\perp +} }
+u_+ 
\right ]
p^2 f_{sm} .
\eqno{(5)}
$$
We can relate the parallel and perpendicular diffusion coefficients to the Bohm diffusion coefficient defined as
$$
D_{Bohm}=\frac {r_g c}{3} =\frac {pc}{3eB_\parallel }
\eqno{(6)}
$$
where $B_\parallel$ is the magnetic field  along the flux tube on a scale much larger than the CR Larmor radius $r_g$.
In standard plasma theories for the diffusion of particles with a scattering time $\tau_{scat}$ longer than the gyration time 
($\omega _g \tau_{scat} >1$ where $\omega _g =ceB _\parallel /p$),
the diffusion coefficient across a magnetic field is $D_\perp= D_{Bohm}/(\omega _g \tau_{scat})$,
and the diffusion coefficient parallel to the magnetic field is
$D_\parallel= D_{Bohm}\times (\omega _g \tau_{scat})$ to within numerical factors of order one that depend
on the nature of the scattering (eg Braginskii 1965, Blandford \& Eichler 1987).
For a parallel shock, the zeroth order uniform parallel magnetic field, and therefore the Bohm diffusion coefficient, is the same upstream and downstream.
Consequently, 
$\sqrt {D_{\parallel -}D_{\perp -} }=\sqrt {D_{\parallel +}D_{\perp +} }= D_{s}$,
where $D_s$ is the Bohm diffusion coefficient near the shock.

As shown in the above equations (3-5), CR acceleration depends on $D_s$ alone, and not on $D_\perp$ and $D_\parallel$ separately. 
This was previously demonstrated by
Eichler (1981) who highlighted the important result that the multiple $D_s^2 =D_\parallel D_\perp$ does not depend on the strength of the turbulence
since the factor $\omega _g \tau _{scat}$ cancels out.
This contrasts favourably with acceleration by an isolated shock where the maximum energy is reduced from
the Hillas energy if the CR mean free path is greater than its Larmor radius and $D_\parallel > D_{Bohm}$,
in which case
$T_{max} \sim (D_\parallel/D_{Bohm})^{-1} u_s Bl$.
In young SNR this is not a major concern since observation and theory both indicate that the CR diffusion coefficient is close to
$D_{Bohm}$ (Stage et al 2006, Uchiyama et al 2007, Bell 2014),
but more generally this need not be so.
Hence a significant advantage of acceleration in a flux tube is that the maximum energy
robustly depends on $D_{Bohm}$ and the maximum CR proton energy is $T_{max} \sim  u_s Bl$.

If  CR with low momenta are injected into the acceleration process uniformly across the width of the shock, $F_{m}=(-1)^m (4/\pi) F_{s}/(2m+1)$
where the spectrum at low momentum  is $F_s p^{-4}$.
For a strong shock with velocity $u_s$, $u_-=u_s$ and $u_ +=u_s/4$. giving
$$
f_{sm}=
(-1)^m \frac {4}{\pi} \frac { F_{s} p^{-4}}{2m+1}
 \left (\frac {1}{2}+  \sqrt { \frac {1}{4}+ \frac {\chi _m ^2}{4} }\ \right ) ^{2}
\left (\frac {1}{2}+  \sqrt { \frac {1}{4}+ 4 \chi _m ^2 }\ \right ) ^{1/2}
\exp \left (
\frac {5}{2} -  
\sqrt { 4+ 4 \chi _m ^2   } -   \sqrt { 1/4+ 4 \chi _m ^2 } 
  \right )
$$
$$
{\rm and } \hskip .5 cm 
\frac {\partial n}{\partial t}= u_sl  \sum _{m=0}^\infty
\frac {(-1)^m}{2m+1}
 \big (
   \sqrt {16 +16 \chi _m ^2 } +  \sqrt {1 +16 \chi _m ^2 } -3
\big )
p^ 2f_{sm}
\hskip .7 cm
{\rm where } \  \
\chi _m=(2m+1)\frac { 2 \pi D_s}{u_sl}\ .
\eqno{(7)}
$$
In equations (7), $f_{sm}$ relates to the CR momentum spectrum at the shock while
$\partial n/\partial t$ relates to 
the momentum spectrum of CR released from the flux tube into the surrounding lobe.
$\partial n/\partial t$ is the crucial quantity if we are interested in the
lobes of radio galaxies as a source of UHECR.  
Note that the spectrum of CR escaping the flux tube is not the same as the spectrum at the shock.
$\partial n/\partial t$ is plotted in figure 3.
The $m=0$ mode makes the major contribution.
The neglect of higher $m$ modes would not make a discernable difference to the curve in figure 3.
\begin{figure}
\includegraphics[angle=0,width=6cm]{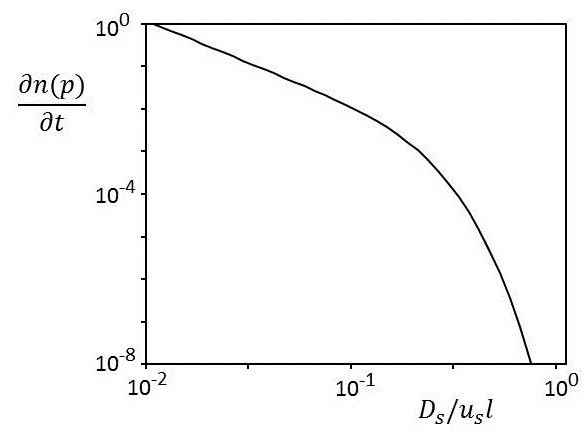}
\centering
\caption{
Normalised differential momentum spectrum $\partial n /\partial t$ (equation 7) of CR produced by a single shock in a flux tube of width $l$.
The horizontal axis is a normalised momentum axis since
$D_s/u_s l$ is proportional to the CR momentum.
}
\label{fig:figure3}
\end{figure}
\noindent
At low CR momentum, the spectrum follows the usual $p^{-2}$ power law for acceleration at a strong shock.
 Figure 3 shows that the momentum spectrum turns over where $D_s/u_sl \approx 0.2$, which corresponds to a maximum
proton energy in eV of
$$
T_{max} \approx 0.6 u_s B l.
\eqno {(8)}
$$
Nuclei with charge $Ze$ can be expected to be accelerated to $Z$ times this energy.
The maximum energy in equation (8)  is close to the Hillas energy defined as $u_s B l$.
$T_{max}$ is independent of the value of $\omega _g \tau _{scat}$.
Our analysis assumes that the entrance to the flux tube is further  from the shock than the diffusion length
$D_{\parallel -}/u_s$ upstream of the shock, and that 
the exit end of the flux tube is further from the shock than the diffusion length $4 D_{\parallel +}/u_s$
downstream of the shock.
If either of these conditions is violated the maximum CR energy is determined by losses from the relevant end of the flux tube.

The reason the maximum CR energy is independent of $\omega _g \tau _{scat}$ is that 
the acceleration rate is proportional to $ D_\parallel ^{-1} \propto  (\omega _g \tau _{scat}) ^{-1}$,
and the loss rate from the sides of the flux tube is proportional to 
$D_\perp$ which is also proportional to $(\omega _g \tau _{scat})^{-1}$.
The maximum CR energy is the energy at which the  loss rate is equal to the acceleration rate.
Since both the acceleration rate and the loss rate are proportional to $(\omega _g \tau _{scat})^{-1}$,
$T_{max}$ is independent of $\omega _g \tau _{scat}$.


 \section{\large\bf Comparison with 2nd-order Fermi acceleration}
\vskip 0.05 cm
Simulations by Matthews et al (2019) followed fluid elements (modelled as tracer particles embedded in the flow)
as they pass along the jet into the hotspot at the head of the jet, and then emerge through a back-flowing flux tube into the radio lobe.
Analysis of the history of tracer particles shows that the flow through the flux tube  is generally transonic with 
supersonic episodes during which shocks occur.
About 10 percent of the tracer particles pass through a shock with a Mach number greater than three
where strong CR acceleration can be expected.
More generally, tracer particles pass through more frequently occuring weak shocks with low Mach numbers.
We could repeat the analysis of the previous section for a multiplicity of weak shocks distributed randomly along the length of the flux tube.
However, a collection of many weak shocks can equivalently be described as
turbulence  with a turbulent velocity
comparable with the large-scale background flow velocity.  
Flow though weak shocks can be treated as a small velocity perturbation
that can be decomposed into Fourier modes as discussed below.
The flow structure in the following analysis is one-dimensional, and therefore does not qualify correctly as
turbulence, but the underlying acceleration process is closely similar to second-order Fermi  acceleration in
a turbulent plasma and for ease of expression we will refer to the disordered velocity field as turbulence.

In this section we analyse CR acceleration by considering second-order Fermi acceleration
in a flux tube with a time-independent spatially-varying flow velocity directed  along the flux tube.  
The configuration is similar to that in figure 2 except that changes in flow velocity $u$ are no longer localised at a single shock
but distributed along the length of the tube.
The flow velocity is assumed to be uniform across the width of the flux tube:
$$
u(z)= u_0 +u_1(z)\ .
\eqno{(9)}
$$
This formulation can be used to represent flow along a flux tube as it passes through a series of constrictions
imposed by the surrounding lobe.
For transonic flow containing weak shocks $u_1 \sim u_0$, but we make the approximation appropriate for
second-order Fermi that $u_1 \ll u_0$ and keep terms of order $(u_1/u_0)^2$.
The implications of increasing the ratio $u_1/u_0$ to $\sim 1$ can then be considered.
In the rest frame of plasma flowing along the flux tube, this looks like waves with amplitude $u_1$ travelling in reverse direction toward
the hotspot at velocity $-u_0$.
The transfer of energy from the turbulence to the CR can be interpreted as transit-time damping in which
CR extract energy from the wave by diffusing out of dense compressed regions on the same timescale as the flow alternates between compression and rarefaction.  This damping of hydrodynamic waves is well known in other contexts (eg Bell 1983).  It is a manifestation of second-order Fermi acceleration since CR preferentially spend time in parts of the flow undergoing compression where they scatter between converging fluid elements.

We apply the free boundary escape condition ($f=0$) at the edges of the flux tube.
The CR distribution can be Fourier expanded,
$f_m \propto \cos ((2m+1) \pi x/ l)$ as in equation (2).
In the case of acceleration by a single shock, the $m=0$ mode is the main source of CR acceleration.
The same applies here, so we develop the analysis by ignoring all modes except $m=0$.
As part of the linearisation procedure, we express the CR distribution function as a sum
of zeroth and first order contributions:
$$
f(z,x,p)=\left [ f_0(p)+f_1(z,p) \right ] \cos (k _\perp x)
\eqno{(10)}
$$
where $k_\perp =\pi/l$. 
$f_0$ and $u_0$ are both uniform along the flux tube.
The analysis proceeds by linearising equation (1) for the CR distribution:
$$
u_0 \frac {\partial f_1}{\partial z} 
- D_\parallel  \frac {\partial ^2 f_1}{\partial z ^2} 
+k_ \perp ^2 D_\perp f_1
=
\frac{1}{3} \frac {\partial u_1}{\partial z} p \frac{\partial f_0}{\partial p}
\eqno {(11)}
$$
where $D_\perp$ \& $D_\parallel$ are uniform both along (in $z$) and across (in $x$) the flux tube.
The solution to equation (11) is
$$
f_1 (z,p)=
\frac{1}{3 u_0 \xi}  p \frac{\partial f_0}{\partial p}
\left [
\int _{-\infty}^z  \frac {\partial u_1}{\partial z'} e^{b(z'-z)} dz' 
+
\int _{z}^\infty  \frac {\partial u_1}{\partial z'} e^{a(z'-z)} dz' 
\right ]
\eqno {(12)}
$$
$$
{\rm where} \hskip 2 cm
a=\frac{u_0 (-1-\xi)} {2D_\parallel}\ , \hskip .5 cm
b=\frac{u_0 (-1+\xi ) } {2D_\parallel}
\hskip .5 cm {\rm and} \hskip .5 cm
\xi =\sqrt {1+ \frac {4k _\perp ^2 D_\perp D_\parallel}{u_0^2}}. \hskip 3 cm
\eqno {(13)}
$$
Equation (12) says that any local gradient in $u_1$, which might be viewed as a weak shock, causes a disturbance in the surrounding CR population
that decays exponentially over a distance $|b|^ {-1}$ upstream and a distance $|a|^{-1}$ downstream, where $a<0$ and $b>0$. 

As in equation (4) of section (2), the rate at which the number of CR with momenta greater than p increases in a  long flux tube
of length $L$,
 $L \gg \max(|a|^{-1},|b|^{-1})$,
is
$$
\frac {\partial N}{\partial t}=
\frac{2l}{\pi}
\int _{0} ^ {L} dz
\int _{p} ^ {\infty} 
\left [
\frac{1}{3} \frac {\partial u_1}{\partial z}  \frac {1}{p^2} \frac{\partial  ( p^3 f_1)}{\partial p}
\right ] 4 \pi p^2 dp
=-
\frac{8l}{3}
\int _0 ^L
 \frac {\partial u_1}{\partial z} p^3 f_1
dz.
\eqno {(14)}
$$
We now express $u_1(z)$ and $f_1(z)$ as Fourier transforms:
$
u_1(z)= (2 \pi)^{-1/2} \int _{-\infty}^\infty u_k(k _\parallel ) e^{ik _\parallel z} dk _\parallel
$
and 
$ f_1(z)= (2 \pi)^{-1/2} \int _{-\infty}^\infty f_k(k _\parallel) e^{ik _\parallel z} d  k_\parallel $.
The application of standard Fourier procedures to equation (11) leads to
$$
f_k(k _\parallel )=\frac {ik_\parallel u_k (k_\parallel)}{3} 
\frac {(D_\parallel k_\parallel^2+D_\perp k_\perp^2) - ik_\parallel u_0}
{ (D_\parallel k_\parallel^2+D_\perp k_\perp^2)^ 2 +k_\parallel^2 u_0^2}\ 
p \frac {\partial f_0} {\partial p}
\eqno {(15)}
$$
$$
{\rm and} \hskip .5 cm
\frac {\partial N}{\partial t}=
- \frac{8 L l }{9 u_0 ^2}
p^4  \frac {\partial f_0} {\partial p} 
\int _{-\infty} ^ \infty 
\Gamma (k_\parallel)
\langle u^2 \rangle_k  dk_\parallel
\hskip .5 cm {\rm where} \hskip .5 cm
\Gamma (k_\parallel)=k_\parallel u_0\left (
\frac {D_\parallel k_\parallel^2+D_\perp k_\perp^2}{ k_\parallel u_0} 
+  \frac { k_\parallel u_0}  {D_\parallel k_\parallel^2+D_\perp k_\perp^2}
\right )^{-1}
\ .
\eqno {(16)}
$$
$\langle u^2 \rangle_k dk _\parallel $ is the mean square velocity of turbulence in the range $k _\parallel $ to $k _\parallel +dk _\parallel $.
$\Gamma (k_\parallel)$ has the dimensions of a rate in time.
$\Gamma (k_\parallel)$ controls the contribution of turbulence at wavenumber $k_\parallel$ to the acceleration of CR with momentum $p$.
At very long wavelengths $\Gamma(k_\parallel)$ is proportional to $k _\parallel ^2$: $\Gamma (k_\parallel) \rightarrow (u_0^2 /D_\perp)(k_\parallel^2/k_\perp ^2)$.
At very short wavelengths $\Gamma(k_\parallel)$ is constant: $\Gamma (k_\parallel) \rightarrow u_0^2/D_\parallel$.
At intermediate wavelengths, there are two critical wavenumbers:
(a) $k_{c \parallel} = u_0 / D_\parallel $, which determines whether the inverse wavenumber $k_\parallel^{-1}$
is shorter ($k_\parallel > k_{c \parallel}$) than the characteristic diffusion length along the flux tube, and
(b) $k_{c\perp}= k_\perp \sqrt {D_\perp/D_\parallel}$ which determines whether parallel diffusion dominates  (for $k_\parallel > k_{c\perp}$)
over perpendicular diffusion.
$\Gamma (k_\parallel)$ can be written in terms of these two critical wavenumbers:
$$
\Gamma (k_\parallel)= \left ( \frac {u_0^2}{D_\parallel} \right) \frac {k_\parallel}{k_{c\parallel}} 
\left ( \frac {k_\parallel^2+k_{c\perp}^2}{k_{c \parallel} k_\parallel} + \frac {k_{c \parallel} k_\parallel} {k_\parallel^2+k_{c\perp}^2} \right )^{-1}
 . \eqno {(17)}
$$
For comparison with shock acceleration, apart from a dimensionless numerical factor, ${u_0^2}/{D_\parallel}$ is the acceleration rate  of CR with
diffusion coefficient $D_\parallel$ at a shock with velocity $u_0$.
The function $\Gamma (k_\parallel )$ is plotted in figure 4.
\begin{figure}
\includegraphics[angle=0,width=7cm]{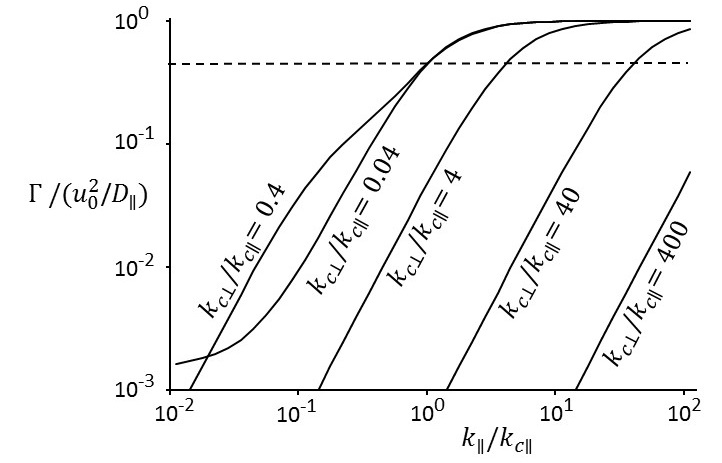}
\centering
\caption{
The function $\Gamma $ in equation (17) for various normalised flux tube widths $l$, where
$l=\pi ( k_{c\perp} / k_{c\parallel} )^{-1} (D_\perp D_\parallel/u_0^2)^{1/2}$.
The dashed line gives the value of $\Gamma$ at which $k=k_{cut}$
as defined in equation (18).
} 
\label{fig:figure4}
\end{figure}
\noindent
If the cut-off wavenumber $k_{cut}$ is defined as the value of $k_\parallel $ at which $\Gamma = 0.5 u_0^2/D_\parallel $, then
$$
k_{cut} = \frac{u_0}{D_\parallel}
 \left [
 \frac {1}{2} +\sqrt{       \frac {1}{4}+ \frac {k_{ \perp}^4 D_{Bohm}^2}{u_0^2 }       }\ 
\right ]^{1/2}
 \eqno{(18)}
$$
where $D_{\perp}D_{\parallel}=D_{Bohm}^2$ and $D_{Bohm}$ is the Bohm diffusion coefficient
as defined in equation (6).
Once again we see that diffusion through the sides of the flux tube is controlled by the multiple of $D_{\perp}$ and $D_{\parallel}$ which is independent
of the strength of the turbulence as represented by $\omega _g \tau _{scat}$.

From equation (16), power-law turbulence  at long wavelengths, $k_\parallel < k_{cut}$, $\langle u^2 \rangle _k \propto k _\parallel ^{-\gamma}$,
does not contribute significantly to CR acceleration
provided $\gamma <2$.
CR either escape through the side of the flux tube before gaining energy, 
or else remain adiabatically fixed in the same fluid element without being accelerated.
On the other hand, at short wavelengths, $k_\parallel> k_{cut}$,  power law turbulence  makes little contribution to CR acceleration provided  $\gamma >0$ 
 since $\langle u^2 \rangle _k \propto k _\parallel ^{-\gamma}$ decreases towards large $k_\parallel$.
Consequently, power law turbulence with $0<\gamma < 2$ contributes most strongly to CR acceleration at wavenumbers
close to $k_{cut}$.
To a reasonable approximation, since $\Gamma (k_\parallel) \approx u_0^2/D_\parallel $ for $k_\parallel > k_{cut}$,
$$
\frac {\partial N}{\partial t}=
-\frac{8Ll}{9 D_\parallel} p^4
  \frac{\partial f_0}{\partial p}
\int _{k_{cut}} ^\infty
\langle u^2 \rangle _k
d k_\parallel .
\eqno {(19)}
$$


The rate $dp/dt$ at which CR gain momentum is related to $\partial N/\partial t$ by the continuity equation:
$
{\partial N}/{\partial t}=Ll\ 4 \pi p^2 f_0 (dp/dt)
$.
The acceleration time for second-order Fermi acceleration time is $\tau _{F2}=p/(dp/dt)$. Hence from equation (19)
$$
 \tau _{F2}=
\frac {9 \pi}{2}  D_\parallel \left (- \frac {p}{f_0} \frac {\partial f_0}{\partial p} \right )^{-1}
\left ( 
\int _{k_{cut}} ^\infty
2 \pi \langle u^2 \rangle _k
d k_\parallel
\right )^{-1}.
\eqno {(20)}
$$
For comparison, the acceleration time for first-order Fermi shock acceleration (Lagage \& Cesarsky 1983a,b) is
$$
\tau _{F1}=\frac {8D_\parallel}{u_s^2}
\eqno {(21)}
$$
where the  representative numerical factor of 8 depends on the ratio of the diffusion coefficients upstream
and downstream of the shock.
Hence, the ratio of the first- and second-order Fermi acceleration times is
$$
\frac{\tau_{F1}}{\tau _{F2}}=\frac {16}{9 \pi}
\left (- \frac {p}{f_0} \frac {\partial f_0}{\partial p} \right )
\left ( \frac {1}{u_s^2}
\int _{k_{cut}} ^\infty
 \langle u^2 \rangle _k
d k_\parallel
\right ).
\eqno{(22)}
$$
The precise value of the factor of $16 /9 \pi$ has limited significance since it is sensitive to assumptions and approximations built into our
derivation. 
More crucially, equation (22) shows, as might be expected, that the rate of second-order Fermi acceleration is of the same order as
that of shock acceleration if the characteristic velocity of the turbulence contributing to CR
acceleration is comparable to the shock velocity.

The acceleration rate is important since the maximum energy reached by CR
is determined by competition between acceleration and loss.
$\tau _{F2} \sim \tau _{F1}$ is required if the second-order Fermi process is
to accelerate CR to the same energy that can be achieved by shock acceleration (equation 8).
That is,  the turbulent flow velocity needs to be comparable
with the velocity at which the plasma flows down the flux tube if second-order is
to compete with first-order Fermi acceleration.
For second-order Fermi acceleration to contribute significantly to
the 	acceleration of UHECR, the turbulence must have 
an energy density comparable to the kinetic energy density of large scale flows in the lobe
as previously indicated by Hardcastle (2010) and consistent with O'Sullivan et al (2009)
who find that wave speeds in excess of $0.1c$ are needed for UHECR acceleration by stochastic Alfven waves  in 
the radio galaxy Centaurus A.

The other main point for comparison between first- and second-order Fermi acceleration is
the slope of the CR  spectrum.
CR escape through the walls of the flux tube is responsible for the turnover in the spectrum
at high energy, but escape through the walls has negligible effect on the spectrum at lower energies.
At energies below the turnover, the spectrum at a single isolated shock 
 is determined by competition
between acceleration at the shock and CR escape 
due to advection downstream away from the shock.
Both the acceleration rate and the advective loss rate are well determined in shock acceleration, 
leading naturally to $n(p) \propto p^{-2}$ below the spectral turnover as in equation (7).
However, the situation is more complicated in the case of multiple weak shocks.
The first order
Fermi spectrum is not simply $n(p) \propto p^{-2}$
since the spectrum is steepened by low compression at weak shocks, yet conversely it is flattened 
by passage through multiple shocks (Bell 1978b).

In the case of second-order Fermi acceleration both the acceleration rate and the loss rate 
can take a range of values, thus leading to a range of possible spectra
depending on competition between the two rates.
The acceleration rate is proportional to the amplitude of the turbulence (equation 20).
The only possible 
escape route for low energy CR is by advection with the fluid flow
through the downstream exit from the flux tube.
If the plasma transit time $L/u_0$ through the flux tube is much longer than the acceleration time, CR loss by advection during acceleration is also  negligible, and  CR number conservation dictates that $\partial N/\partial t$ is the same at all momenta.
From equation (19), this leads to a CR momentum  spectrum determined by
$$
\frac {p^4}{D_\parallel} \frac {\partial f_0}{\partial p}
  \int _{k_{cut}} ^\infty
 \langle u^2 \rangle _k
d k_\parallel = {\rm constant}.
\eqno {(23)}
$$
According to equation (23) a range of CR spectra are possible in this steady state, depending on the wavenumber spectrum of the turbulence.
Turbulence in a flux tube is a special case,
but if the turbulence follows a Kolmogorov power law with energy density proportional to $k  ^{-5/3}$,
such that $\int _{k _{cut}}^\infty  \langle u^2 \rangle _k  dk_\parallel \propto p^{2/3}$,
since  $k_{cut}\propto D_\parallel ^{-1} \propto p^{-1}$,
then $f_0(p) \propto p^{-8/3}$ and $n(p) \propto p^{-2/3}$.
This would be an unusually flat CR spectrum with most of the CR energy density at the  large momentum end of the spectrum.
The cause of the flat spectrum is that CR acceleration is slower at high CR momentum,
with the result that a relatively greater number of CR are needed at high momentum to maintain a constant flux of CR 
from low to high momentum.
The momentum spectrum given by equation (23) is unlikely except in special circumstances, 
and an extended  calculation is needed that takes account of time dependence, escape through the ends of the flux tube,  and feedback processes across the momentum spectrum.

\section{Conclusions}

We examine the acceleration of CR in hydromagnetic flux tubes with application to the lobes of radio galaxies
where flux tubes contain shocks that are mostly low Mach number (Matthews et al 2019).
We consider CR acceleration in the framework of two overlapping formalisms: first-order Fermi  acceleration by shocks and second-order Fermi acceleration by strong turbulence. 
We show that the maximum CR energy at a single shock in a flux tube is close to the Hillas energy $ZuBl$
where $u$ is the flow velocity along the flux tube, $B$ is an ordered field aligned along the flux tube,
$l$ is the width of the flux tube,
and the CR charge is $Ze$.
In contrast to CR acceleration at the outer shocks of SNR,
the attainment of the Hillas energy is not dependent on Bohm diffusion in
the shock environment.
Consequently, the Hillas energy is a relatively robust estimate of the energy to which CR are accelerated.

We show that second-order Fermi acceleration in turbulence is slower than first-order shock acceleration
by the ratio of the mean square turbulent velocity to the square of the shock velocity.
Hence, second-order acceleration in weak turbulence is less able to compete with CR losses,
and cannot accelerate CR to the highest energies.
However, if the turbulence is strong in the sense that the mean square turbulent velocity is comparable 
with the large scale flow velocity, then second-order acceleration competes well with first-order acceleration
since it is broadly equivalent to multiple 
first-order acceleration episodes as a CR passes through a succession of weak shocks.

The flow within lobes of radio galaxies is more complicated (see references in section 1) than allowed by our simplifed model of one-dimensional flow 
in a steady uniform straight flux tube, and additional acceleration processes may be active.
For example, the strong velocity differential across the edges of flux tubes may be a prime location for shear acceleration 
(Earl et al 1988, Ostrowski 1998, Rieger \& Duffy 2006, Caprioli 2015) 
and for the growth of strong multi-dimensional turbulence due to the Kelvin-Helmholtz instability.  
Also, variations in the width of flux tubes along their length may give rise to lateral compressions and rarefactions that respectively accelerate and decelerate CR, although resulting changes in CR energy may be approximately adiabatic with limited overall effect. 
As shown by Achterberg (1981),
second-order Fermi acceleration is an umbrella term for a number of different 
effects that can be combined in a general quasi-linear theory of wave-particle interactions.
The more general theory may provide a fruitful basis for an extended analysis in the complicated environment of a flux tube, and
it may be possible to encompass first- and second-order Fermi acceleration and shear acceleration in a single comprehensive analysis (Lemoine 2019).

The model we use here could be extended to include the above additional effects,
but within its limitations, our 
analysis supports the  proposal by Matthews et al (2018, 2019) that UHECR
may be accelerated in the lobes of radio galaxies.
Suitably extended, our analysis may  be applicable to other steady confined flows such as jets.

\section {ACKNOWLEDGEMENTS}
We thank Bram Achterberg for helpful comments. 
This research was supported by the UK
Science and Technology Facilities Council under grant No. ST/N000919/1.

\section{REFERENCES}
Achterberg A., 1981, A\&A, 97, 259
\newline
Araudo A.T., Bell A.R., Blundell K.M., 2015, ApJ, 806, 304
\newline
Araudo A.T., Bell A.R., Blundell K.M., Matthews J.H., 2018, MNRAS, 473, 3500
\newline
Axford W.I., Leer E., Skadron G., 1977,  Proc 15th Int. Cosmic Ray Conf., 11, 132
\newline
Bell A.R., 1978a, MNRAS, 182, 147 
\newline
Bell A.R., 1978b, MNRAS, 182, 443
\newline
Bell A.R., 1983, Phys Fluids, 26, 269
\newline
Bell AR, 2014, Braz J Phys, 44, 415
\newline
Bell A.R.,  Araudo A.T., Matthews J.H., Blundell K.M., 2018, MNRAS, 473, 2364
\newline
Bell A.R., Schure K.M., Reville B., Giacinti G.,2013,  MNRAS, 431, 415
\newline
Blandford R.D.,  Eichler D., 1987,  Phys Rep, 154,1
\newline
Blandford R.D.,  Ostriker J.P.,  1978, ApJ, 221, L29
\newline
Braginskii S.I., 1965, Rev Plasma Phys, 1, 205
\newline
Caprioli D., 2015, ApJ, 811, L38
\newline
Cielo, S., Antonuccio-Delogu, V., Macciò, A. V., Romeo, A. D., Silk, J., 2014, MNRAS, 439, 2903
\newline
Earl J.A., Jokipii J.R., Morfill G., 1988, ApJ, 331, L91
\newline
Eichler D., 1981, ApJ, 244, 711
\newline
Eichmann, B., Rachen, J. P., Merten, L., van Vliet, A., Becker Tjus, J., 2018, JCAP02, 036
\newline
Falle S.A.E.G., 1991,  MNRAS,  250, 581
\newline
Fermi E., 1949, Phys Rev, 75, 8
\newline
Hardcastle M.J., 2010, MNRAS, 405, 2810
\newline
Hardcastle M.J., Cheung C.C., Feain I.J., Stawarz L., 2009, MNRAS, 393, 1041
\newline
Hillas A.M., 1984, ARA\&A, 22, 425
\newline
Jones F.C., 1994, ApJSS, 90, 561
\newline
Jokipii J.R., 1982, ApJ, 255, 716
\newline
Jokipii J.R., 1987, ApJ, 313, 842
\newline
Keppens R., Meliani Z., van der Holst B., Casse F., 2008, A\&A, 483, 663
\newline
Kirk J.G.,  Reville B., 2010,  ApJL, 710, L16
\newline
Krause, M., 2005, A\&A, 431, 45
\newline
Krymskii G.F.,  1977, Sov Phys Dokl, 23, 327
\newline
Lagage P.O., Cesarsky C.J., 1983a, A\&A, 118, 223
\newline
Lagage P.O., Cesarsky C.J., 1983b,  ApJ, 125, 249
\newline
Lee M.A. 1982, J Geophys Res, 87, 5063
\newline
Lemoine M., 2019, Phys Rev D, 99, 083006
\newline
Lemoine M., Pelettier G., 2010,  MNRAS, 402, 321
\newline
Matthews J.H., Bell A.R., Blundell K.M., Araudo A.T., 2018, MNRAS, 479, L76
\newline
Matthews J.H., Bell A.R., Blundell K.M., Araudo A.T., 2019, MNRAS, 482, 4303
\newline
Mignone, A., Rossi, P., Bodo, G., Ferrari, A., Massaglia, S., 2010, MNRAS, 402, 7
\newline
Norman M.L. Winkler K.-H., Smarr L., Smith M.D., 1982, A\&A, 113, 285
\newline
Ostrowski M, 1998, A\&A, 335, 134
\newline
O'Sullivan S., Reville B., Taylor A.M., 2009, MNRAS, 400, 248
\newline
Reiger F.M., Duffy P., 2006, ApJ, 652, 1044
\newline
Reville B., Bell A.R., 2014, MNRAS, 439, 2050
\newline
Saxton C.J., Sutherland R.S., Bicknell G.V., Blanchet G., Wagner S.J., Metchnik M.V., 2002, A\&A, 393, 765
\newline
Scheuer P.A.G.S., 1995, MNRAS, 277, 331 
\newline
Stage M.D., Allen G.E., Houck J.C. Davis J.E., 2006, Nature Physics, 2, 614
\newline
Tchekhovskoy A., Bromberg, O., 2016, MNRAS, 461, 46
\newline
Uchiyama Y. et al, 2007, Nature, 449, 576
\newline
Webb G.M., Bogdan T.J., Lee M. A., Lerche I., 1985,  MNRAS, 215, 341
\newline
Wykes S., Croston J.H., Hardcastle M.J., Eilek J.A., Biermann P.L.,
Achterberg A., Bray J.D., Lazarian A., Haverkorn M., Protheroe R.J., Bromberg O.,
2013, A\&A, 558, A19
\newline
Zirakashvili V.N., Ptuskin V.S., 2008,  ApJ, 678, 939
\newline\vskip 0.25 cm

\noindent{\bf APPENDIX A}
\newline
In this Appendix we find a solution to equation (1) by decomposing the distribution function $f$ into Fourier modes as defined in equation (2).  Since $\partial u /\partial z =0$ everywhere except at $z=0$, 
the equation is linear in each of the upstream and downstream plasmas separately, with 
exponential scaleheights $k_{m\pm}$ in $z$ defined by
$$
k_{m\pm} u_\pm- k_{m\pm}^2 D_{\parallel \pm} + k_{ m } ^2 D_{\perp \pm} =0
\eqno {(A1)}
$$
where  $k_{ m} =(2m+1)\pi/l$, and $u_{\pm}$, $D_{\parallel \pm}$ \& $D_{\perp \pm}$ are the upstream and downstream
values of $u$, $D_{\parallel}$ \& $D_{\perp }$.
The solution of equation (A1) is
$$
k_{m\pm } D _{\parallel \pm}= \frac{1}{2} \left (  u_{\pm }  \mp  \sqrt {u_{\pm } ^2 +4 k_m^2  D_{\parallel \pm} D_{\perp \pm} } \right ) .
\eqno{(A2)}
$$
$f$ is continuous across the shock $f_s(p,x)=f_-(0,x,p)=f_+(0,x,p)$.
Integrating equation (1) in $z$ across the shock gives
$$
 \frac{1}{3p^2 f_{ms}} \frac{\partial (p^3 f_{ms})}{\partial p}=
1 -   \frac {k_{m+} D_{\parallel +}- k_{m-} D_{\parallel -}}{  u_+  -u_-}
\eqno{(A3)}$$
where $f_{ms}$ is the value of $f_m$ at the shock ($z=0$).
Substituting equation (A2) into the equation (A3) gives
$$
 \frac{p}{3 f_{sm}} \frac{\partial  f_{sm} }{\partial p}=
-\frac{1}{2}
- \frac {    \sqrt {u_-  ^2 +4 k_m^2  D_{\parallel -} D_{\perp -} } +  \sqrt {u_+  ^2 +4 k_m^2  D_{\parallel +} D_{\perp +} } }
{2(  u_- -u_+)} .
\eqno{(A4)}
$$
In the limit of a strong shock ($u_-=4u_+$) in a wide flux tube ($k_m=0$), the solution is $f_s \propto p^{-4}$, which is the standard power law solution for diffusive shock acceleration.

We assume that the CR scattering length is proportional to the CR Larmor radius,
in which case the multiple of $D_\perp$ and $D_\parallel$ is proportional to $p^2$.
Using the standard integral
$$
\int \sqrt {u^2+b^2 p^2}\  \frac{dp}{p}= \sqrt {u^2+b^2 p^2} + u \log(p)- u \log (u+ \sqrt {u^2+b^2 p^2})+{\rm constant}
\eqno{(A5)}
$$
equation (A4) can be integrated to give
$$
f_{sm}=F_{m} p^{-3u_ - /(u_--u_+)} 
\left [1+  \sqrt { 1+\frac {4 k_m ^2 D_{\parallel -}D_{\perp -} } {u_-^2} }\ \right ] 
^{3u_-/2(u_--u_+)}
\left [1+  \sqrt { 1+\frac {4 k_m ^2D_{\parallel +}D_{\perp +}  } {u_+^2} }\ \right ] 
^{3u_+/2(u_--u_+)}
$$
$$
\times
\exp \left [ -  \frac{3}{2(u_--u_+)} \left (
\sqrt { u_-^2 + 4 k_m ^2 D_{\parallel -}D_{\perp -}   }
+  \sqrt { u_+^2 + 4 k_m ^2 D_{\parallel +}D_{\perp +}  }  \right )  \right ]
\eqno{(A6)}
$$
where $F_m$ is a constant of integration.

\end{document}